# Theory of deeply subwavelength electromagnetic structuring of semiconductors due to thermal-ionization instability

*I. V. Oladyshkin*, D. I. Kulshin, E. O. Smolina*

*A.V. Gaponov-Grekhov Institute of Applied Physics of the Russian Academy of Sciences, 46 Ul'yanov Street, Nizhny Novgorod, 603951, Russia*

*oladyshkin@ipfran.ru

We develop a theory of ionization-assisted instability that governs the reflection of intense electromagnetic pulses from a semiconductor surface. Local overheating of the material increases the free-carrier density and conductivity, which in turn enhances the absorption of the incident radiation and further amplifies local heating and avalanche ionization. At high pump repetition rates, the overheating accumulates from pulse to pulse, leading to a gradual rise of the local material temperature and free carrier density – a positive feedback loop that drives thermal runaway. However, this effect competes with diffraction of the pump field on the emerging perturbation of the dielectric permittivity or on the surface relief, which reduces the local electric field amplitude in the overheated regions, thus providing negative feedback. We show that the competition between heat diffusion and diffraction near-field weakening creates conditions for the exponential growth of temperature perturbations with a certain wavenumber determined by the pulse repetition rate, pump field, thermal conductivity and other material parameters. The derived analytical expressions accurately describe the spatial period of deeply subwavelength structures observed in recent experiments on silicon structuring by intense terahertz radiation, and reproduce qualitatively dependencies of their period on the pump intensity, the incident wavelength and the number of pump pulses.

## INTRODUCTION

Laser-induced periodic surface structures (LIPSS), spontaneously formed upon irradiation of a solid surface by intense laser pulses, have attracted considerable research attention for over half a century after the first observation of this phenomenon by M. Birnbaum [1]. The possibility of contactless and maskless surface modification with spatial resolution exceeding the diffraction limit opens broad prospects for LIPSS applications in photonics and optoelectronics, as well as in fabrication of hydrophobic or antireflective coatings.

The laser-induced structuring in itself is a clear manifestation of complex physical processes taking place in the material under intense irradiation – excitation of new electromagnetic modes, development of instabilities, or nonlinear feedback between the electromagnetic field and the medium response. Despite decades of research, these processes often remain unexplained, which means that the corresponding regimes of laser-matter interactions are not fully understood.

Classical models of LIPSS formation on the surface of metals, dating back to the works of J. Sipe [2, 3] and developed in subsequent studies, relate the emergence of periodic relief to the interference of incident radiation with surface waves – surface plasmon-polaritons (SPPs). Within these models, the spatial period of the structures is equal to SPP wavelength (and close to the pump wavelength $\lambda_0$), while the orientation of the stripes is perpendicular to the polarization of the incident radiation [2, 3]. These predictions are in good agreement with numerous experiments in the visible and near-infrared ranges.

However, recent experimental results [4, 5] do not fit into the framework of classical models. In Refs. [4, 5], it was demonstrated that irradiation of silicon substrates by intense terahertz (THz) and mid-infrared (mid-IR) pulses leads to the formation of periodic structures with a spatial period as small as 1/25 to 1/10 of the pump wavelength, with the stripes oriented parallel to the polarization of the pump electric field. Such a small period and orientation of the surface relief cannot be explained either by interference of electromagnetic waves at the pump frequency or by linear conversion of the pump radiation into other spatial modes, including SPPs, since the predicted periods turn out to be an order of magnitude larger than those observed. These results deserve particular attention because they point to a principal possibility of deeply subwavelength surface structuring – that is, of overcoming the diffraction limit by a significant margin – and thus call for a physical mechanism capable of explaining such a strong subwavelength localization. To the best of our knowledge, no physical mechanism has yet been proposed to explain the results from Refs. [4, 5].

In the present work, we develop a theory of formation of deeply subwavelength LIPSS, based on the ionization-assisted instability upon reflection of intense THz and mid-IR pulses from a semiconductor surface. We show that the key factor is the competition between two processes: (1) avalanche impact ionization, leading to a local increase of the free-carrier density and enhancement of radiation absorption in overheated regions, and (2) diffraction of the pump radiation on the perturbation of the dielectric permittivity, which weakens the electric field in regions with enhanced ionization.

The physical picture of the process is as follows (see Fig. 1). Intense incident radiation ionizes the near-surface layer of the semiconductor [6–8], creating a thin conductive layer (electron-hole plasma). Any spatial perturbation of the electron density leads to modulation of the dielectric permittivity, on which the incident radiation undergoes diffraction. The diffracted fields provide negative feedback: in regions with higher electron concentration the field is weakened. However, the lattice heating caused by Joule absorption and subsequent impact ionization create positive feedback, amplifying the initial perturbation.

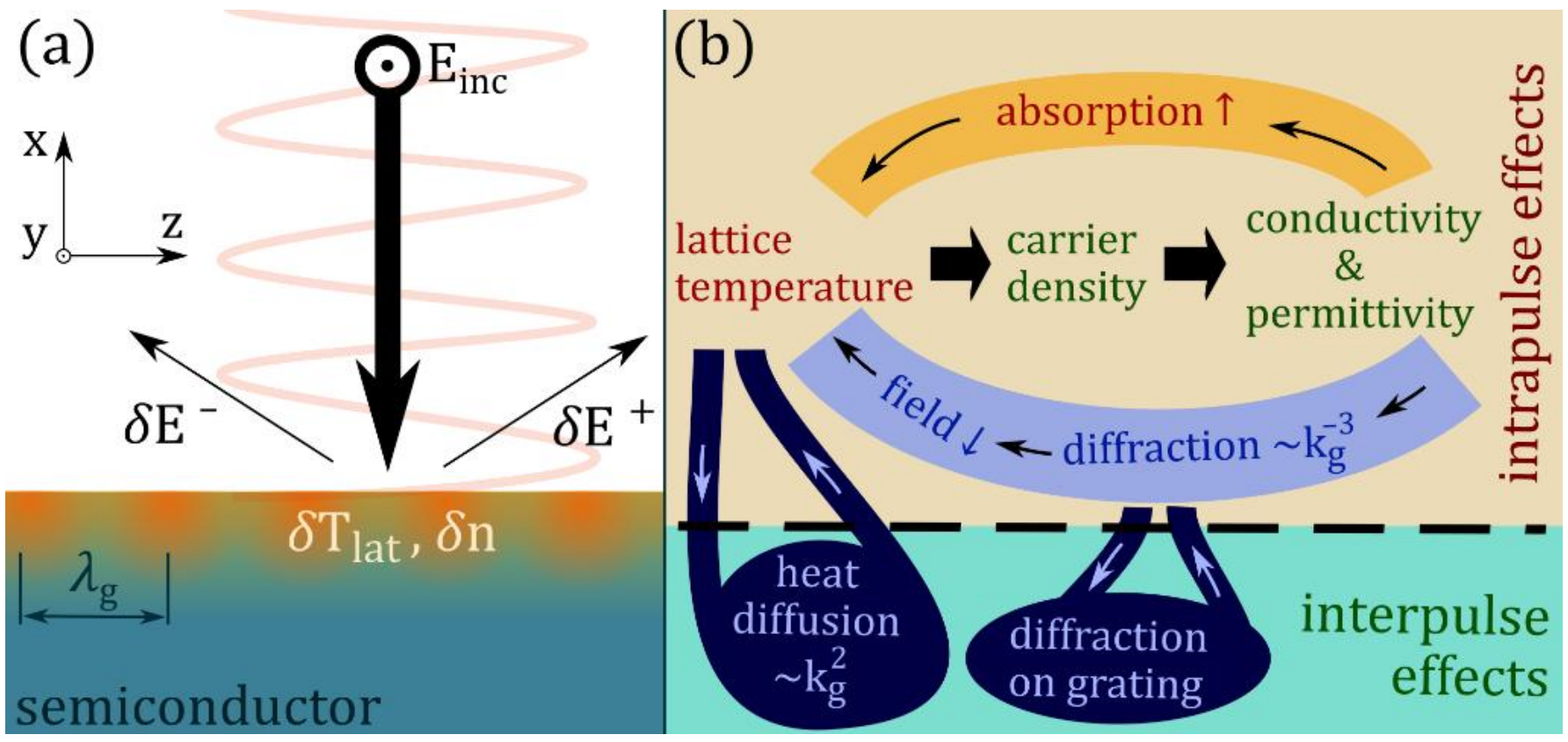

Figure 1. (a) Geometry of the problem: a linearly polarized electromagnetic wave, which is incident normally on a semiconductor surface and diffracts ($\delta E^{\pm}$) and the periodic modulation of the medium permittivity, created by the carrier density modulation. $\lambda_g$ is the spatial period of the modulation (the LIPSS period). (b) Physical picture of the thermal-ionization instability. The upper yellow loop shows the positive feedback: local overheating increases the carrier density and conductivity, which enhances the absorption and further increases the heating. The blue loops show the negative feedback: diffraction of the pump field on the permittivity modulation and on the surface corrugation (Section IV) weakens the field in the overheated regions, while thermal diffusion tends to smooth out the temperature perturbation between subsequent pump pulses.

At high pump repetition rates (27 MHz and 2.86 GHz in experiments [4, 5]), the heating accumulates from pulse to pulse, being smoothened by thermal diffusion between pulses. The competition between thermal instability and diffraction-induced field weakening leads to the emergence of an optimal perturbation wavelength at which the instability growth rate is maximal. Below, combining well-established models of carrier dynamics in semiconductors with the standard diffraction problem, we obtain an analytical expression for the instability increment and calculate the optimal spatial period of LIPSS. The resulting expressions appear to explain deeply subwavelength structuring of silicon by terahertz pulses [4] and the nonlinear dependence of the LIPSS period on the pump intensity and frequency.

We then extend the analysis to the case of a corrugated surface, which becomes relevant at later stages of the structuring process, when the relief produced by the preceding pulses is no longer negligible. We show that the additional diffraction on the surface corrugation modifies the resonance condition and shifts the optimal perturbation wavenumber towards higher values, i.e., towards shorter LIPSS periods. This provides a natural explanation for the experimentally observed decrease of the LIPSS period with increasing number of pump pulses [4], and for the difference between the periods measured in the center and at the periphery of the irradiated spot, where the local fluence – and hence the accumulated relief depth – is different. Despite the simplifying assumptions made to keep the model transparent and to arrive at expressions suitable for analysis and prediction, the developed theory demonstrates good quantitative agreement with experiment. The derived expressions for the optimal LIPSS period are robust with respect to variations of poorly constrained material parameters.

The paper is organized as follows. In Section I, as a necessary preliminary step, we solve the problem of diffraction of an electromagnetic wave on a periodic perturbation of the dielectric permittivity near the semiconductor surface and obtain expressions for the diffracted fields in the regimes of weak and strong ionization. Section II is devoted to the derivation of coupled equations for heating and ionization of the medium during a single pump pulse. In Section III, we pass to the “slow” evolution of the temperature perturbations under the action of many pump pulses, find the instability growth rate and its optimum with respect to the spatial period, comparing theoretical predictions with experimental data [4]. In Section IV, we analyze the role of the corrugated surface and show how the growth of the relief depth shifts the optimal perturbation wavenumber and leads to a gradual decrease of the LIPSS period, in qualitative agreement with the experimental observations. Finally, in Section V we discuss the limits of applicability of the theory, using the experiment of Ref. [5] as an example, where the inter-pulse delay is comparable to the Auger recombination time and the material modification is accompanied by the generation of structural defects.

## I. DIFFRACTION ON THE PERMITTIVITY PERTURBATIONS

The starting point for the subsequent analysis is the diffraction of the pump radiation on a perturbation of the medium dielectric permittivity. In this paper we focus on low-frequency pumping; accordingly, the conductivity is governed solely by intraband electron motion, which we describe within the Drude model. Interband conductivity becomes relevant only at photon energies on the order of the bandgap energy, which is about 1 eV in silicon and even higher in many other semiconductors and dielectrics. Below we consider THz and mid-IR pumping (aiming to interpret experimental results reported in Ref. [4] and, partially, in Ref. [5]), so the photon energy is below 0.1 eV.

In what follows we use the term "diffraction" in its broad sense, which includes both propagating and evanescent (non-propagating) diffraction orders. In the deeply subwavelength regime considered here, $k_g \gg k_0\sqrt{|\bar{\varepsilon}|}$, all diffracted orders are evanescent, so that the diffracted field remains localized near the surface and does not carry energy away from it.

Consider a quasi-monochromatic electromagnetic wave at the frequency $\omega$ polarized along $y$ axis incident normally on the semiconductor surface (the boundary plane is $x = 0$, see Fig. 1). Evolution of the electric field inside the medium can be described by Maxwell equations

$$rot\,\mathbf{H} = \frac{1}{c}\frac{\partial \varepsilon(\mathbf{r})\mathbf{E}}{\partial t}, \tag{1}$$

$$rot\,\mathbf{E} = -\frac{1}{c}\frac{\partial \mathbf{H}}{\partial t}, \tag{2}$$

where $\mathbf{E}$ and $\mathbf{H}$ are electric and magnetic field, respectively, $\varepsilon(\mathbf{r})$ is the medium permittivity (describing the contributions of both bound and free electrons) and $c$ is the speed of light. Supposing that the medium permittivity is homogeneous along $y$ axis, so that $\varepsilon(\mathbf{r}) = \varepsilon(x, z)$, we reduce the problem to the scalar wave equation on $E_y$ component:

$$\Delta E_y + \varepsilon(x, z)k_0^2 E_y = 0, \tag{3}$$

where $k_0 = \omega/c$ is the pump wavenumber in vacuum. Further we introduce time-dependent permittivity $\varepsilon(t)$, and the Eq. (3) will be valid as long as the variation of $\varepsilon$ are adiabatically slow in comparison with the frequency $\omega$.

We use the Drude model of the semiconductor permittivity

$$\varepsilon = \varepsilon_0 - \frac{\omega_p^2}{\omega(\omega + i\nu)}, \tag{4}$$

where $\varepsilon_0$ is the background permittivity ($\varepsilon_0 \cong 11.7$ for undoped Si in low-frequency range), $\omega_p$ and $\nu$ are the plasma frequency and the collision rate of free charge carriers, respectively. The plasma frequency $\omega_p$ is given by

$$\omega_p = \sqrt{\frac{4\pi n e^2}{m^*}}, \tag{5}$$

where $n$ is the free carrier density and $m^*$ is their effective mass. To take into account contributions of free electrons and holes, we define $m^*$ as $m^* = \sqrt{\frac{m_e m_h}{m_e + m_h}}$, where $m_e$ and $m_h$ are the effective masses of electrons and holes, respectively. In what follows, we decompose the carrier density and hence the plasma frequency into a uniform background $n_0, \omega_p$ and a spatially varying perturbations $\delta n(x,z), \delta\omega_p(x,z)$, treating the latter as a small correction.

The zero-order solution of Eq. (3) for the electric field inside the semiconductor (without spatial perturbations of plasma frequency) can be written as follows:

$$E_y(x,t) = E_y^0 \exp(\eta x - i k_x x - i\omega t), \tag{6}$$

where

$$\eta = \pm\, k_0\, Im \sqrt{\varepsilon_0 - \frac{\omega_p^2}{\omega(\omega + i\nu)}}, \tag{7}$$

$$k_x = \pm\, k_0\, Re \sqrt{\varepsilon_0 - \frac{\omega_p^2}{\omega(\omega + i\nu)}}. \tag{8}$$

Here, the signs must be chosen so that $\eta > 0$ (exponential decay into the sample in the $-x$ direction) and $k_x > 0$ (propagation away from the boundary into the bulk).

To find the first-order perturbation of the electric field $\delta E_y$ (i.e. the amplitude of diffraction waves), we write the medium permittivity as

$$\varepsilon = \bar{\varepsilon} + \delta\varepsilon(x,z), \tag{9}$$

where $\bar{\varepsilon} = \varepsilon_0 - \frac{\omega_p^2}{\omega(\omega+i\nu)}$ and $\delta\varepsilon(x,z) = -\frac{\delta\omega_p^2(x,z)}{\omega(\omega+i\nu)}$ are the homogeneous part and the spatial perturbation of the permittivity, respectively. So that, the transmitted wave amplitude $E_y^0$ (zero-order solution) is determined by the Fresnel equation $E_y^0 = 2E_y^{inc}/\left(\sqrt{\bar{\varepsilon}} + 1\right)$, where $E_y^{inc}$ is the incident wave amplitude. Also, since here we focus on high-intensity regimes [4, 5], corresponding to pump electric fields of the order of 1–3 MV/cm, the collision frequency $\nu$ supposed to be saturated at the value of $\sim 10^{15} s^{-1}$ [8–10] and so has no significant variations in the interaction region.

In the first perturbation order, Eq. (3) has the following form:

$$\Delta\delta E_y + k_0^2\left(\bar{\varepsilon}\,\delta E_y + E_y \delta\varepsilon\right) = 0. \tag{10}$$

Consider a particular spatial harmonic of the dielectric permittivity perturbation $\delta\varepsilon$, having the wavenumber $k_g$ along $z$ axis and localized in the near-surface layer of the depth $\gamma^{-1}$:

$$\delta\varepsilon(x,z) = \delta\varepsilon_0 e^{\gamma x}\cos k_g z\,. \tag{11}$$

Further we relate the dynamics of the perturbation $\delta\varepsilon$ to the amplitude of diffracted waves. Seeking the diffracted field in the form

$$\delta E_y(x,z) = \delta E_y(x)\cos k_g z\,, \tag{12}$$

we pass to the equation on its spatial profile inside the semiconductor $\delta E_y(x)$:

$$\frac{\partial^2 \delta E_y(x)}{\partial x^2} + \left(k_0^2\bar{\varepsilon} - k_g^2\right)\delta E_y(x) = -k_0^2\delta\varepsilon_0 E_y^0 e^{-ik_x x+\gamma x}. \tag{13}$$

Eq. (13) is an inhomogeneous differential equation, whose general solution is the following:

$$\delta E_y(x,z) = \left(Ae^{\eta_2 x} - \frac{k_0^2\delta\varepsilon_0 E_y^0}{(\eta - ik_x+\gamma)^2 + \left(k_0^2\bar{\varepsilon} - k_g^2\right)}e^{-ik_x x+\gamma x}\right)e^{-i\omega t}\cos k_g z\,, \tag{14}$$

where $\eta_2 = \pm\sqrt{-k_0^2\bar{\varepsilon} + k_g^2}$ and the sign must be chosen so that $Re\,\eta_2 > 0$ and $Im\,\eta_2 < 0$, similarly the signs in Eqs. (7) and (8).

The arbitrary constant $A$ is determined by the boundary conditions at the semiconductor–vacuum interface. Let us write the diffracted field in vacuum $\delta E_y^v(x,z)$ as

$$\delta E_y^v(x,z) = Ce^{\eta_1 x}\cos k_g z\,, \tag{15}$$

where $C$ is another arbitrary constant and $\eta_1 = \pm\sqrt{k_g^2 - k_0^2}$ ($Re\,\eta_1 < 0$ and $Im\,\eta_1 > 0$). Applying the continuity condition of the tangential electric and magnetic fields at the boundary ($E_y = const$ and $H_z = \frac{1}{ik_0}\frac{\partial E_y}{\partial x} = const$), we find:

$$A = \frac{k_0^2\delta\varepsilon_0 E_y^0}{(\eta - ik_x+\gamma)^2 + \left(k_0^2\varepsilon_0 - k_g^2\right)}\frac{(\eta_1 - \eta + ik_x - \gamma)}{(\eta_1 - \eta_2)}, \tag{16}$$

$$C = \frac{k_0^2\delta\varepsilon_0 E_y^0}{(\eta - ik_x+\gamma)^2 + \left(k_0^2\varepsilon_0 - k_g^2\right)}\left(\frac{\eta_2 - \eta + ik_x - \gamma}{\eta_1 - \eta_2}\right). \tag{17}$$

Taking into account the definition of $\eta_2$, we can further simplify Eq. (17):

$$C = \frac{k_0^2\delta\varepsilon_0 E_y^0}{\eta - ik_x + \gamma + \eta_2}\left(\frac{1}{\eta_2 - \eta_1}\right). \tag{18}$$

Now let us find the amplitude of the total electric field $E_y^{tot}$ near the semiconductor boundary $x = 0$, summarizing the transmitted zero-order field and the diffraction waves:

$$E_y^{tot} = E_y^0\left(1 + \frac{k_0^2\delta\varepsilon_0}{\eta - ik_x + \gamma + \eta_2}\left(\frac{1}{\eta_2 - \eta_1}\right)\cos k_g z\right). \tag{19}$$

Since below we consider heating and avalanche ionization of the medium, which is determined by the field magnitude (rather than its complex amplitude), we take the absolute value of this expression in the first perturbation order:

$$\left|E_y^{tot}\right| \cong \left|E_y^0\right|\left(1 + Re\left[\frac{k_0^2\delta\varepsilon_0}{\eta - ik_x + \gamma + \eta_2}\left(\frac{1}{\eta_2 - \eta_1}\right)\right]\cos k_g z\right). \tag{20}$$

Correspondingly, the total electric field perturbation $\delta|E_y^{tot}|$ has the following form:

$$\delta|E_y^{tot}| \cong |E_y^0| Re\left[\frac{k_0^2\delta\varepsilon_0}{\eta - ik_x + \gamma + \eta_2}\left(\frac{1}{\eta_2 - \eta_1}\right)\right]\cos k_g z\,. \qquad (21)$$

We now turn to several limiting cases that are important for the following analysis. If the medium ionization is relatively weak ($\omega_p \ll \omega, \nu$) and the perturbation wavenumber is small ($k_g \ll k_0$), the propagation constants are given by $\eta \cong 0$, $k_x \cong k_0\sqrt{\varepsilon_0}$, $\eta_2 \cong -ik_0\sqrt{\varepsilon_0}$ and $\eta_1 = ik_0$. Also, the diffraction waves and the transmitted zero-order wave propagate into the medium almost without absorption ($Re[\eta_2] \cong 0, \eta \cong 0$), so that the perturbation of the dielectric permittivity, driven by these fields, exhibits negligible decay into the bulk and $\gamma \cong 0$. In this case Eq. (21) gives

$$\delta|E_y^{tot}| \cong -\frac{|E_y^0|}{2\sqrt{\varepsilon_0}\left(\sqrt{\varepsilon_0}+1\right)} Re(\delta\varepsilon_0)\cos k_g z\,. \qquad (22)$$

Substituting the explicit form of the Drude permittivity (4), we find

$$\delta|E_y^{tot}| \cong \frac{|E_y^0|}{2\sqrt{\varepsilon_0}\left(\sqrt{\varepsilon_0}+1\right)}\frac{\delta\omega_p^2}{(\omega^2+\nu^2)}\cos k_g z\,. \qquad (23)$$

Expressions (22) and (23) are nothing other than the perturbation of the standard Fresnel formula, showing that a homogeneous (i.e., $k_g \to 0$ and $\gamma \to 0$) decrease of the permittivity due to ionization leads to an increase of the transmitted field, as expected. In this limit, the perturbation of the field magnitude does not depend on $k_g$.

If the medium ionization is still weak ($\omega_p \ll \omega, \nu$), but the perturbation wavenumber is high ($k_g \gg \sqrt{\varepsilon_0}k_0$), Eq. (21) transforms into

$$\delta|E_y^{tot}| \cong -|E_y^0|\frac{k_0^2}{4k_g^2}\frac{\delta\omega_p^2}{(\omega^2+\nu^2)}\cos k_g z\,. \qquad (24)$$

So that, the contribution of the diffraction waves is negative and its absolute value decreases as $1/k_g^2$. By the "negative contribution of diffraction" we mean that in regions of enhanced electron density – and hence increased plasma frequency $\delta\omega_p^2$ – the total electric field is diminished by the diffracted field.

In the regime of strong medium ionization, i.e. $\frac{\omega_p^2}{\omega\nu} \gg \varepsilon_0$ (which is most relevant below), the medium is similar to a metal from the electrodynamical point of view, because its permittivity (4) has negative real part and large imaginary part. Both the zero-order transmitted field and the diffraction waves decay exponentially in the skin-layer, which determines the characteristic depth of the permittivity perturbation $\gamma^{-1}$. For the case of small wavenumbers $k_g^2 \ll k_0^2\frac{\omega_p^2}{\omega\nu}$ we find:

$$\delta|E_y^{tot}| \cong -\frac{3}{10}|E_y^0|\frac{\delta\omega_p^2}{\omega_p^2}\cos k_g z\,, \qquad (25)$$

which describes the overall decrease of the transmitted field due to the plasma frequency growth above its critical value. Similarly to Eq. (23), there is no dependence on $k_g$ in the limit of homogeneous perturbation $k_g \to 0$.

Finally, at high $k_g$ we obtain

$$\delta|E_y^{tot}| \cong -|E_y^0|\frac{k_0^2}{6k_g^2}\frac{\delta\omega_p^2}{\nu^2}\left(1+\frac{k_0\omega_p}{3k_g\omega}\sqrt{\frac{\nu}{2\omega}}\right)\cos k_g z\,. \qquad (26)$$

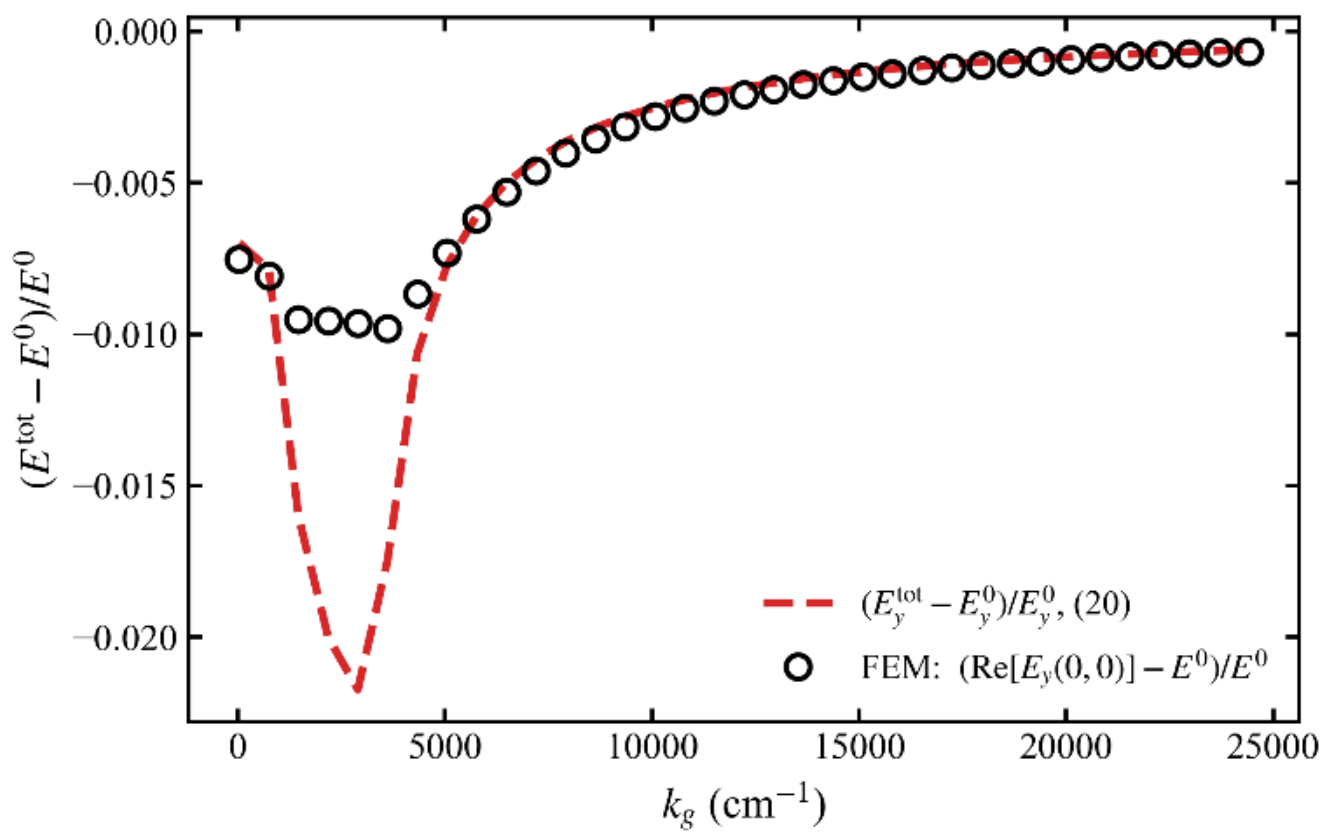

Figure 2. Normalized perturbation of electric field near the semiconductor surface, caused by diffraction, versus the perturbation wavenumber (the case of strong background medium ionization $\omega_p^2/\omega\nu \gg \varepsilon_0$, $k_0 = 10^3\ cm^{-1}$). The result of finite-element modelling of the diffraction problem is shown by circles, and the analytical curve given by Eq. (20) is shown by dashed red line.

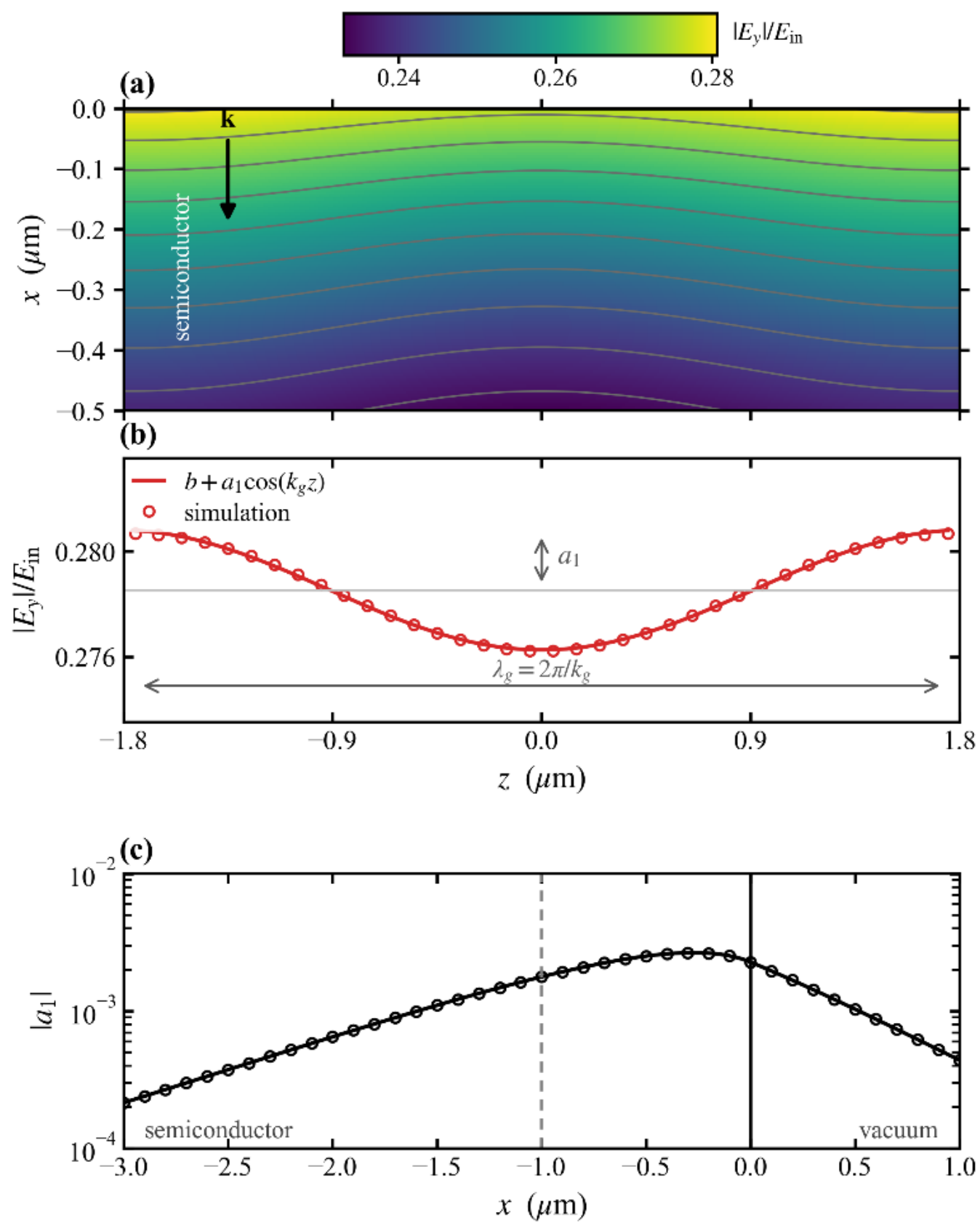

Figure 3. Electric field near the semiconductor surface, normalized to the incident amplitude $E_{inc}$. The results of numerical modelling for the case of strong medium ionization at $k_g = 1.75 \times 10^4\ cm^{-1}$ are shown. (a) Map of the out-of-plane field component $|E_y^{tot}|/E_{inc}$ over one grating period $\lambda_g = 2\pi/k_g$ along the $z$ axis (below the semiconductor interface); the semiconductor-vacuum interface is the plane $x = 0$. (b) Normalized field $|E_y^{tot}|/E_{inc}$ along one grating period at the interface $x = 0$: circles represent the values calculated using FEM, and the curve shows the harmonic fit $|E_y^{tot}|/E_{inc} = b + a_1\cos(k_g z)$. (c) Spatial decay of the field modulation in vacuum and in the medium (along $x$-axis).

Similarly to Eq. (24), the contribution of the diffraction field here is negative and its magnitude decreases with $k_g$. For clarity, here we assumed that $\nu^2 \gg \omega^2$, which corresponds to typical experimental parameters of [4] (where $\nu$ is expected to saturate at $\sim 10^{15}\ s^{-1}$, while the incident frequency is $\sim 10^{13}\ s^{-1}$).

To validate the analytical description of the diffraction problem, we solve it numerically using the finite element method (FEM) for a weakly modulated permittivity layer localized near the surface of a semiconductor and illuminated by an $s$-polarized (TE) wave at normal incidence (see Figs. 2 and 3). As previously, the semiconductor occupies the half-plane $x < 0$; the free space occupies $x > 0$, and $z$ denotes the coordinate along the surface. Owing to the periodicity of the perturbation in $z$ with the wavenumber $k_g$, the computational domain is restricted to a single cell of length $2\pi/k_g$, whose lateral boundaries are connected by the Floquet periodicity condition with the wave vector (0, 0). In the regions far from the surface, the domain is truncated by perfectly matched layers (PML) that absorb all outgoing radiation both in the vacuum and in the bulk of the semiconductor. The incident radiation is a plane wave at normal incidence with the frequency 5 THz, which corresponds to the free-space wavelength $\lambda \approx 60$ µm. All parameters correspond to undoped silicon; the depth of the ionized layer is taken to be $\gamma^{-1} = 1$ µm, and the background permittivity is $\varepsilon_0 = 11.7$.

The semiconductor is assumed to be strongly ionized near the surface, so that the unperturbed medium is no longer homogeneous. In contrast to our simplified analytical approach, in the numerical modelling we take into account explicitly that the background ionization is localized inside the skin-layer and, consequently, that the background permittivity $\bar{\varepsilon}$ is inhomogeneous, see Eq. (9). Thus, the weak permittivity modulation $\delta\varepsilon_0 e^{\gamma x} \cos k_g z$ is imposed on top of the permittivity of the ionized skin layer with $\bar{\varepsilon} = \varepsilon_0 + \varepsilon_p e^{\gamma x}$ (as an example, in Figs. 2 and 3 we choose $\varepsilon_0$= 11.7, the plasma contribution $\varepsilon_p = -30 + 50i$, and the perturbation amplitude $\delta\varepsilon_0 = -1 + i$).

The results of the numerical modelling are compared with the analytical prediction of Eq. (20) in Fig. 2, where the normalized perturbation of the field magnitude, $\delta|E_y^{tot}|/|E_y^0|$, is plotted as a function of the perturbation wavenumber $k_g$. The numerical data confirm the two key features of the analytical solution: (i) the diffraction contribution is negative, i.e., the total field in the regions of enhanced ionization is reduced, and (ii) its magnitude decreases monotonically at high $k_g$. Moreover, for $k_g \gtrsim 5 \times 10^3\ cm^{-1}$ the analytical expressions (20) or (21) reproduce the numerical result with high accuracy.

The deviation observed at lower wavenumbers, $k_g \sim k_0$, is attributed to the finite depth of the ionized layer taken into account in the numerical model. At high perturbation wavenumbers, when the penetration depth of the diffracted waves inside the medium is determined solely by $k_g$ and does not depend on the medium permittivity ($\eta_2 \cong k_g$), the analytical approximation becomes essentially exact. This is the regime we will rely on in what follows.

Below we analyze competition between the diffraction weakening (Eq. (26)) and thermal effects, which are the absorption increase in the regions with high electron density and heat diffusion. Throughout the following sections, we base our numerical estimates on the parameters of silicon, both because the experiments of interest [4, 5] were performed on this material and because its properties are well characterized experimentally. Naturally, this does not restrict the applicability of the developed theory to other semiconductors.

## II. IONIZATION AND HEATING OF THE MEDIUM

Irradiation of a semiconductor by an electromagnetic wave with the amplitude of the order of 1 MV/cm leads to very fast heating of its near-surface layer and impact carrier multiplication. The dominant ionization mechanism in semiconductors under the action of strong non-resonant radiation is the avalanche (impact) processes, which are much more efficient than multiphoton or tunneling ionization [6–11].

Avalanche ionization takes place when the external electric field directly accelerates carriers up to kinetic energies, comparable to the bandgap, during the free pass time. Also, significant carriers' heating can further assist ionization due to effective decrease of the bandgap for hot carriers [12]. The basic description of the field-driven avalanche ionization is given by the Chynoweth's model [11]:

$$\frac{dn}{dt} = n v_d \alpha_0 \exp\left(-\frac{b}{\left|E_y^{tot}\right|}\right), \tag{27}$$

where $\alpha_0 e^{-b/\left|E_y^{tot}\right|}$ is the ionization probability per unit electron path length, $b$ is the characteristic threshold field and $v_d$ is the drift electron velocity. For pump pulses of picosecond duration, the Auger recombination term $\Gamma n^3$ (where $\Gamma$ is the Auger relaxation constant) can be neglected in Eq. (27), see estimations below Eq. (32). In general, parameters $\alpha_0$, $b$ and $\Gamma$ also depend on the electric field and, overall, on the state of the ionized material, so they can be considered as constants only within a certain electric field range. Basing on experimental data for silicon [7, 8, 13], we can rely on the following values for estimates: $\alpha_0 \cong (3-5) \times 10^5\ cm^{-1}$, $b \cong 1.2 \times 10^6\ V/cm$ and $\Gamma \cong 2 \times 10^{-31}\ cm^3/s$.

According to works [8–10], in electric fields of the amplitude of hundreds of kV/cm and higher the effective collision frequency in silicon increases up to $10^{15}\ s^{-1}$, while the drift velocity $v_d$ saturates at $\sim 10^7\ cm/s$. Also, the effective mass of overheated carriers approaches the electron mass in vacuum [14, 15]. From above it follows that the avalanche ionization rate in silicon at high pump field $E \approx b$ reaches the value of $(3-5) \times 10^{12}\ s^{-1}$. In comparison with it, the Auger relaxation rate $\Gamma n^2$ remains negligibly small for the electron densities up to $\sim 10^{21}\ cm^{-3}$, which corresponds to the plasma frequency of $2.5 \times 10^{15}\ rad/s$. It means that, for low-frequency pumping, the main limiting factor for the avalanche ionization development is the incident wave screening by free carriers, rather than Auger processes.

Let us consider the case of strong medium ionization, which is more relevant for the experimental conditions of Ref. [4]. In this work, the pump pulse duration was 20 ps, the field amplitude was 2–3 MV/cm and the pump frequency varied from 3 to 6 THz. If the incident wave screening by the electron-hole plasma is not taken into account, these parameters give an unrealistic multiplication factor of charge carriers of $\sim 10^{20}$.

Below we use the simplest model of the incident radiation screening, assuming that the transmitted wave magnitude (zero-order field in terms of Section I) can be described by the Fresnel formula for a homogeneous medium with the permittivity $\bar{\varepsilon}$:

$$\left|E_y^0\right| \cong \frac{2E_{inc}}{\sqrt{|\bar{\varepsilon}|}}, \tag{28}$$

where

$$\bar{\varepsilon} \cong \frac{i\omega_p^2}{\omega\nu}, \tag{29}$$

since $\omega_p^2 \gg \omega\nu$ and $\nu \gg \omega$ in the THz range (which also means that $|\bar{\varepsilon}| \gg 1$). This permittivity model corresponds to the standard Drude model of a metal with the low-frequency conductivity $\sigma = \frac{ne^2}{m^*\nu}$.

Note that more precise model of pump transmission inside the medium should be based on the solution of nonlinear wave equation, where the increase rate of the local electron-hole density $dn(x,t)/dt$ is determined by the local electric field $\left|E_y^0(x,t)\right|$, according to Eq. (27). However, under the experimental conditions [4, 5], the accumulated heating of the semiconductor surface up to the melting temperature creates the overcritical background density of free electrons of about $10^{18} - 10^{19}\ cm^{-3}$ [16] even in thermal equilibrium. As a result, the medium becomes strongly metallic already at the background level, which partially justifies the use of the simple screening model of Eqs. (28)–(29), which captures the essential physics of the field penetration.

Under these assumptions, Eq. (27) transforms into

$$\frac{dn}{dt} = nv_d\alpha_0 \exp\left(-\frac{b}{E_{inc}}\sqrt{\frac{n\pi e^2}{m^*\omega\nu}}\right), \tag{30}$$

which can be rewritten in the dimensionless form:

$$\frac{d\rho}{dt} = a\rho\exp(-\rho), \tag{31}$$

where $\rho = \frac{b}{E_{inc}}\sqrt{\frac{\pi n e^2}{m^*\omega\nu}}$ and $a = \frac{v_d\alpha_0}{2}$. The exact solution of Eq. (31) can be expressed implicitly through the exponential integral function $Ei(\rho)$. For relatively long pump pulses (as in Ref. [4]) we can use the asymptotic limit $at = v_d\alpha_0 t/2 \gg 1$, which gives $\rho(t) \cong \ln(at)$ or, in the original physical notations,

$$n(t) = \frac{E_{inc}^2}{b^2}\frac{m^*\omega\nu}{\pi e^2}\left(\ln\left(\frac{v_d\alpha_0 t}{2}\right)\right)^2. \tag{32}$$

So that, in the regime of strong ionization, the density of free carriers demonstrates slow logarithmic increase with time. For the incident field magnitude $E_{inc} = 2 - 3\ MV/cm$ (estimated pump field used in [4]), the incident frequency of 5 THz and for the parameters of silicon listed above we obtain the electron density of $(0.5 - 2) \times 10^{21}\ cm^{-3}$ at $t = 20\ ps$, which corresponds to the plasma frequency of $(1.5 - 3) \times 10^{15}\ rad/s$ and the skin layer thickness of $\sim$1 μm. This supports our initial assumption that the medium is metal-like ($\omega_p^2 \gg \omega\nu$), while the Auger recombination rate $\Gamma n^2 \approx 10^{10} - 10^{11} s^{-1}$ is indeed negligible at the timescales of the pump, except the case of highest field magnitude. Finally, the dynamics of the background plasma frequency can be directly expressed from Eq. (32):

$$\omega_p(t) = \sqrt{\omega\nu} \times \frac{2E_{inc}}{b}\ln\left(\frac{v_d\alpha_0 t}{2}\right), \tag{33}$$

which also describes the growth of the permittivity absolute value $|\bar{\varepsilon}| \cong \omega_p^2/\omega\nu$ and decrease of the electric field inside the semiconductor as $\left|E_y^0\right| \cong b/\ln\left(\frac{v_d\alpha_0 t}{2}\right)$. Note again that Eqs. (32) and (33) are valid only for $v_d\alpha_0 t \gg 1$.

Now we turn to analysis of the perturbation dynamics, taking place on the background of the homogeneous medium ionization. Based on the results of Section I, here we focus on relatively high perturbation wavenumbers $k_g$, since in this region the negative contribution of the diffraction field is minimized and, consequently, we expect the maximal instability increment at high $k_g$.

Consider a density perturbation $\widetilde{\delta n}(t)$ with a wavenumber $k_g$:

$$\widetilde{\delta n}(t) = \delta n(t)e^{\gamma x}\cos k_g z, \tag{34}$$

which evolution is described by the first perturbation order of Eq. (27):

$$\frac{d\delta n}{dt} = \delta n v_d\alpha_0 \exp\left(-\frac{b}{\left|E_y^0\right|}\right) + nv_d\alpha_0 \exp\left(-\frac{b}{\left|E_y^0\right|}\right)\frac{b\delta\left|E_y^{tot}\right|}{\left|E_y^0\right|^2}. \tag{35}$$

Using Eq. (26) for the electric field decrease due to interference with the diffraction fields, we find:

$$\frac{1}{\delta n}\frac{d\delta n}{dt} = \frac{1}{n}\frac{dn}{dt}\left(1 - \frac{k_0^2}{12k_g^2}\frac{b}{E_{inc}}\frac{\omega_p^2}{\nu^2}\sqrt{\frac{\omega_p^2}{\omega\nu}}\left(1 + \frac{\omega_p k_0}{3\omega k_g}\sqrt{\frac{\nu}{2\omega}}\right)\right). \tag{36}$$

Eq. (36) can be integrated in a general case; however, to keep further expressions compact, we restrict ourselves to the limit $\frac{1}{3ck_g}\sqrt{\frac{\omega_p^2\nu}{2\omega}} \gg 1$, which will be validated below. In this case we obtain a simple analytical relation between the perturbation $\delta n(t)$ and the background density $n(t)$:

$$\delta n(t) \cong \delta n_0 \frac{n(t)}{n_0} \exp\left[-\frac{1}{72\sqrt{2}}\frac{k_0^3}{k_g^3}\frac{b}{E_{inc}}\frac{\left(\omega_p^4(t)-\omega_{p,0}^4\right)}{\omega^2\nu^2}\right], \qquad (37)$$

where $\delta n_0$, $n_0$ and $\omega_{p,0}$ are the carriers' density perturbation, the background density and the plasma frequency at the moment $t = 0$ (i.e., in the moment of the pump arrival), respectively. Eq. (37) can be further simplified by neglecting $\omega_{p,0}^4$ relatively to $\omega_p^4$ (since the initial carrier density is much smaller than once reached after a few picoseconds) and by linearization of the exponent.

The values $\delta n_0$ and $n_0$ can be determined as the thermally equilibrium densities of charge carriers for the local temperature of the medium at the moment of the pump pulse arrival [16]:

$$n_0 = n_{cv} \exp\left[-\frac{W_g}{2kT_{lat}}\right], \qquad (38)$$

$$\delta n_0 = \frac{W_g \delta T_{lat}}{2kT_{lat}^2} n_0, \qquad (39)$$

where $n_{cv}$ is the effective density of states of the conduction and valence bands, $W_g$ is bandgap energy, $T_{lat}$ is the crystal lattice temperature (equal to the electronic temperature before the interaction) and $k$ is the Boltzmann constant. Writing expression (38), we assume that the medium is localy thermalized during the delay time between subsequent pump pulses, which is 37 ns in the experiment [4]. Also remind that the material constants $n_{cv}$ and $W_g$ are temperature-dependent, which, however, is not critical for the subsequent analysis.

To determine the medium temperature and its perturbations, let us start with the equation on the energy balance equation for free electrons during a single pump pulse:

$$\frac{d}{dt}\left(C_e T_e + W_g n\right) = eEv_d n - C_e G(T_e - T_{lat}), \qquad (40)$$

where $T_e$ and $C_e = \frac{3}{2}kn$ are the electron gas temperature and heat capacity per unit volume, respectively, and $G$ is the electron-lattice thermalization rate. The term $eEv_d n$ represents the power delivered by the electric field to the electrons, with the drift velocity $v_d$ taken at its saturation value. We neglect heat diffusion in Eq. (40), since the diffusion length per the pump duration (20 ps in [4]) is only few tens of nanometers, which is much smaller than the skin layer depth (~1 μm). Heat diffusion is taken into account at inter-pulse timescales in Section III, where it plays an essential role.

Since the thermalization time is typically a few hundred femtoseconds ($G \approx 5 \times 10^{12}\ s^{-1}$ in Si [17]), we can pass to the quasi-stationary limit of Eq. (40):

$$eEv_d n \cong C_e G(T_e - T_{lat}). \qquad (41)$$

This yields the temperature difference $(T_e - T_{lat})$ at which the electronic temperature is saturated and all the absorbed energy is transferred to the crystal lattice. So that, in this regime we obtain:

$$C_{lat}\frac{dT_{lat}}{dt} = eEv_d n, \qquad (42)$$

where $C_{lat}$ is the crystal lattice heat capacity per unit volume.

The first perturbation order of Eq. (42) with Eq. (26) gives[1]:

$$C_{lat}\frac{d\delta T_{lat}}{dt}=\delta n\times ev_d\left|E_y^0\right|\left(1-\frac{1}{18\sqrt{2}}\left(\frac{\omega_p}{\sqrt{\nu\omega}}\frac{k_0}{k_g}\right)^3\right). \tag{43}$$

Substituting $\delta n$ from Eqs. (37)–(39) and linearizing the result, we find:

$$\frac{d\delta T_{lat}}{dt}=\delta T_{lat}^0\frac{W_g v_d eE_{inc}}{3k^2T_{lat}^2}\frac{\omega_p\sqrt{\nu\omega}}{\omega_a^2}\left(1-\frac{1}{18\sqrt{2}}\frac{k_0^3}{k_g^3}\left(\frac{\omega_p}{\sqrt{\nu\omega}}\right)^3\left(1+\frac{b}{4E_{inc}}\frac{\omega_p}{\sqrt{\nu\omega}}\right)\right). \tag{44}$$

Here we have taken the high-temperature (or Dulong–Petit) limit for the lattice heat capacity $C_{lat}=3kn_a$, where $n_a$ is the atomic density, and introduced the constant $\omega_a^2=\frac{4\pi n_a e^2}{m^*}$. The only time-dependent value in Eq. (44) is the plasma frequency $\omega_p(t)$, which demonstrates slow logarithmic increase (see Eq. (33)). So that, Eq. (44) can be integrated over the pump pulse duration either exactly or approximately, by replacing the logarithm with its characteristic value.

For the parameters of Si in strong electric fields ($v_d=10^7\ cm/s$, $W_g\cong 0.7\ eV$, $\nu=10^{15}\ s^{-1}$, $m^*=0.5\ m_0$ and $\omega_a=1.8\times 10^{16}\ rad/s$) and typical experimental parameters of the experiment [4] ($E_{inc}\cong 2-3\ MV/cm$, $\omega=5\ THz\cong 3\times 10^{13}\ rad/s$, $kT_{lat}\cong 0.15\ eV$) we obtain that the relative temperature perturbation growth per unit pump pulse $\delta T_{lat}/\delta T_{lat}^0$ is about 5–10 times. At the same time, the estimated background heating of the crystal lattice per single pulse is 500–1000 K (see Eq. (42)), which coincides with the observed sample melting at maximal pump fluences [4]. Because of that, here and below we chose the set of silicon parameters corresponding to its melting temperature. In the next section we formulate the time-averaged equations which describe competition of the thermal runaway with the heat diffusion between pump pulses and predict the optimal value of $k_g$ for the instability development.

### III. AVERAGED EQUATIONS AND INSTABILITY

The interaction of the semiconductor with the train of pump pulses naturally separates into two distinct timescales: short heating episodes during each pump pulse (picoseconds) and long inter-pulse intervals (nanoseconds), in which the only process is heat relaxation due to thermal conductivity. In terms of Eq. (44), the inter-pulse feedback is realized through $\delta T_{lat}^0$ – the residual perturbation of the medium temperature by the time the next pump pulse arrives.

Aiming to describe the heating dynamics under multiple pump pulses, we introduce an averaged heating rate $\overline{\frac{d\delta T_{lat}}{dt}}$ in the following form:

$$\overline{\frac{d\delta T_{lat}}{dt}}=\frac{\delta T_{lat,n+1}-\delta T_{lat,n}}{\tau_d}, \tag{45}$$

where $\delta T_{lat,n}$ and $\delta T_{lat,n+1}$ are the temperature perturbations at the moment of the $n$-th pump pulse arrival and after its end, respectively, and $\tau_d$ is the time delay between pulses. This allows us to write a time-averaged equation on the temperature perturbation growth at long timescales as $\overline{d\delta T_{lat}/dt}=\Omega(k_g)\overline{\delta T_{lat}}$, where $\Omega(k_g)$ is the increment of the temperature perturbation increase:

[1]In the case of TM-polarization (when the stripes are perpendicular to the electric field), the deep-subwavelength response is quasi-electrostatic, and therefore obeys $D_\perp=\varepsilon E_\perp=$ const. Thus, an increase of the local permittivity reduces the local electric field. In the strongly conducting low-frequency limit $\bar{\varepsilon}\cong\frac{4\pi i\sigma}{\omega}$ (see Eq. (29)), the absorbed power per unit volume is $P_{abs}=eEv_d n\propto\bar{\varepsilon}E=const$. Hence, this polarization does not provide the positive heating–absorption feedback underlying the instability discussed here.

$$\Omega(k_g) = \frac{\tau_p}{\tau_d}\frac{W_g v_d e E_{inc}}{3k^2 T_{lat}^2}\frac{\omega_p\sqrt{\nu\omega}}{\omega_a^2}\left(1 - \frac{1}{18\sqrt{2}}\frac{k_0^3}{k_g^3}\frac{\omega_p^3}{(\nu\omega)^{3/2}}\left(1 + \frac{b}{4E_{inc}}\frac{\omega_p}{\sqrt{\nu\omega}}\right)\right) - D_{lat}\left(k_g^2 + \frac{1}{L^2}\right). \quad (46)$$

Here $D_{lat}$ is the heat diffusion coefficient, $L$ is the heating depth (comparable with the skin-layer) and $\tau_p$ is the pump pulse duration. In Eq. (46), the plasma frequency $\omega_p$ should be treated as averaged over the duration of a single pump pulse. Because $\omega_p$ grows only logarithmically in time, this averaging is well justified: in what follows, we may simply set $t \simeq \tau_p$ in Eq. (33) for estimates.

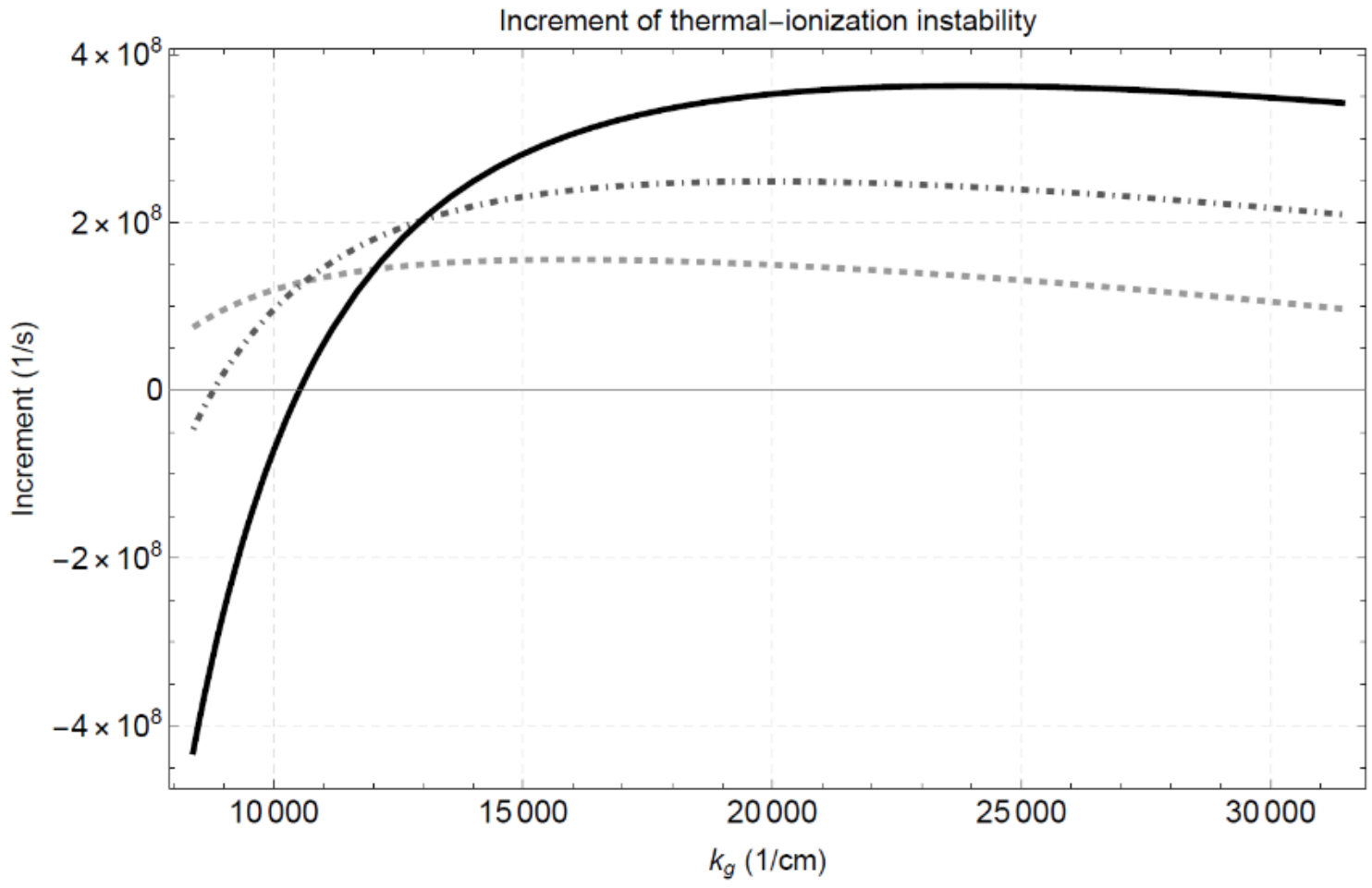

Figure 4. Calculated dependence of the thermal-ionization instability increment Ω (Eq. (46)) on the perturbation wavenumber $k_g$ for silicon at melting temperature ($kT_{lat} = 0.15\,eV$, $v_d = 1.4 \times 10^7 cm/s$, $\nu = 10^{15} s^{-1}$, $b = 1.2\,MV/cm$, $\alpha_0 = 4 \times 10^5 cm^{-1}$, $W_g = 0.7\,eV$, $D_{lat} = 0.1\,cm^2/s$) and for the set of experimental parameters from Ref. [4] ($\tau_p = 20\,ps$, $\tau_d = 37\,ns$, $\omega = 5\,THz$). Solid, dot-dashed and dashed curves correspond to the incident field amplitudes of 3 MV/cm, 2.5 MV/cm and 2 MV/cm, respectively (which are the estimated amplitudes of micro-pulses with the fluences of 0.3, 0.2 and 0.12 J/cm$^2$ in Ref. [4], respectively).

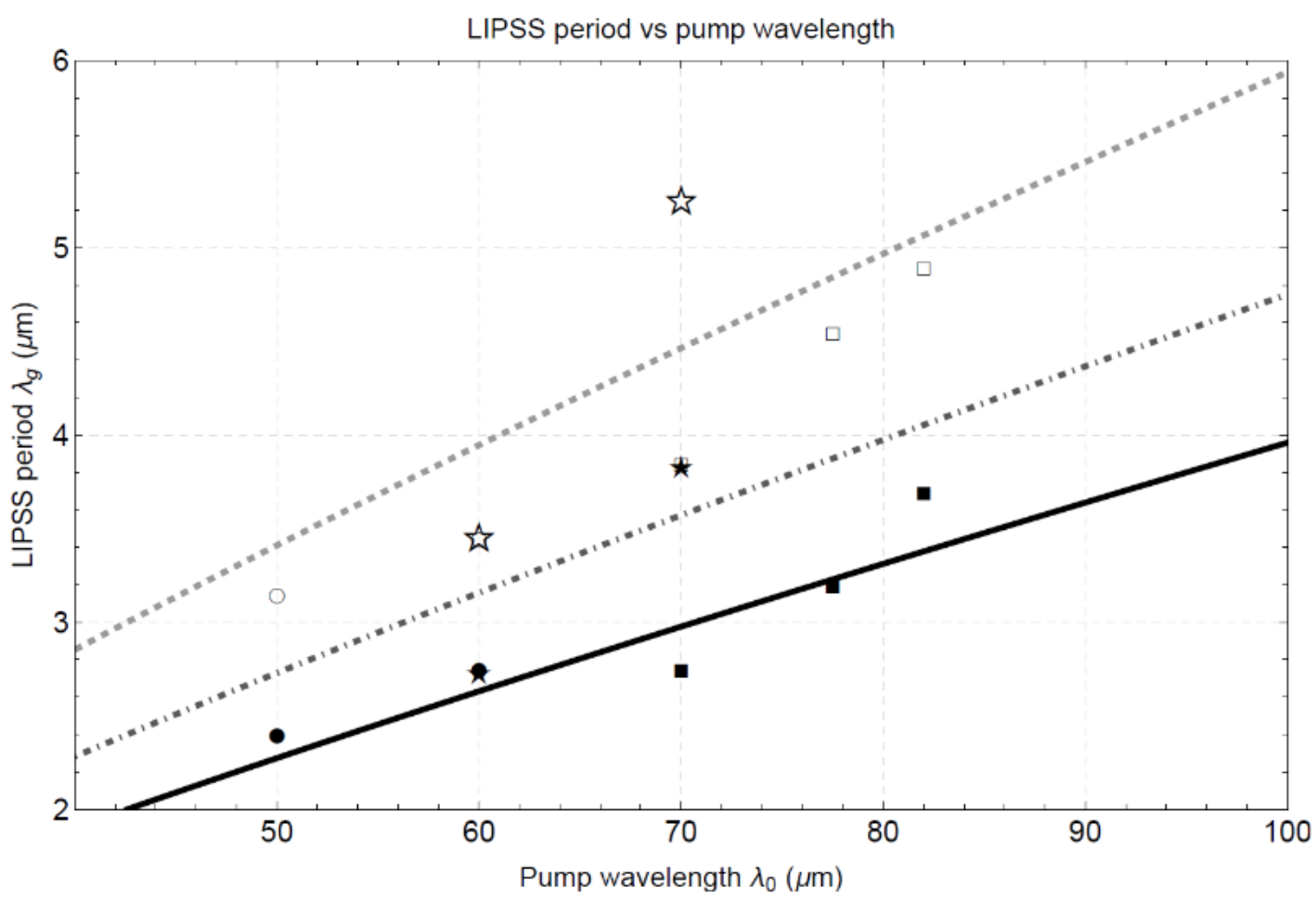

Figure 5. Predicted dependence of LIPSS period on the incident wavelength (Eq. (47)). The set of parameters is the same as in Fig. 4; solid, dot-dashed and dashed curves also correspond to the incident field amplitudes of 3 MV/cm, 2.5 MV/cm and 2 MV/cm, respectively. Squares, stars, and circles are the experimental points for the micro-pulse fluences of 0.3, 0.2 and 0.12 J/cm$^2$ (or estimated field amplitudes of 3 MV/cm, 2.5 MV/cm and 2 MV/cm), respectively [adopted from Fig. 3 (a), Ref. [4]]. Filled symbols denote the center of the irradiated spot; open symbols denote the periphery.

Theoretical dependencies $\Omega(k_g)$ are shown on Figure 4 for the parameters of silicon near the melting temperature and the experimental parameters used in Ref. [4]. As Fig. 4 demonstrates, the increment remains positive in a rather wide range of wavenumbers and reaches maximum at $k_g = (1.5 - 2) \times 10^4 cm^{-1}$, i.e., at spatial periods of 3–4 μm, in good agreement with the experiment.

Calculating the optimal perturbation wavenumber as $\frac{d\Omega(k_g)}{dk_g} = 0$, we find:

$$k_g^* = \left(\frac{k_0^3}{D_{lat}\tau_d}\right)^{1/5} \times \left(\frac{W_g v_d e E_{inc}\tau_p}{36\sqrt{2}k^2 T_{lat}^2}\frac{\omega_p^4}{\nu\omega\omega_a^2}\left(1 + \frac{b}{4E_{inc}}\frac{\omega_p}{\sqrt{\nu\omega}}\right)\right)^{1/5}. \quad (47)$$

Here, the first term has the dimension of the wavenumber, while the others are dimensionless factors. Note that the value of $\sqrt{D_{lat}\tau_d}$ is the thermal diffusion length over the inter-pulse delay (~2 μm per 37 ns), therefore, the optimal $k_g$ is set by both the pump wavenumber and the inverse diffusion length. Substituting the plasma frequency from Eq. (33) and taking into account that $\ln(v_d\alpha_0\tau_p/2) \gtrsim 1$ for simplicity, we finally obtain:

$$k_g^* \cong \left(\frac{k_0^3}{D_{lat}\tau_d}\right)^{1/5} \left(\frac{\sqrt{2}}{9}\frac{W_g v_d eb\tau_p}{k^2 T_{lat}^2}\frac{\nu\omega}{\omega_a^2}\right)^{1/5} \times \frac{E_{inc}}{b}\ln\left(\frac{v_d\alpha_0\tau_p}{2}\right). \quad (48)$$

As an example, calculations for the set of parameters listed after Eq. (44), $D_{lat} \approx 0.1\, cm^2/s$ (the heat diffusivity of silicon near the melting point [18]) and $\tau_d = 37\, ns$ give $k_g^* \cong 2 \times 10^4\, cm^{-1}$, and so the calculated LIPSS spatial period $\lambda_g$ is about 3 μm, which is in very good agreement with the experimental result $\lambda_g \cong 2.7$ μm for the frequency of 5 THz [4]. This also validates the assumption $\frac{1}{3ck_g}\sqrt{\frac{\omega_p^2\nu}{2\omega}} \gg 1$ made while passing from Eq. (36) to Eq. (37).

Figure 5 shows the analytical dependencies of the optimal perturbation spatial period on the incident wavelength for the field amplitudes 2 MV/cm, 2.5 MV/cm and 3 MV/cm (plotted according to Eq. (47)). These values are the estimated amplitudes, corresponding to the micro-pulse fluences of 0.12, 0.2 and 0.3 J/cm$^2$ used in [4]. Experimental points, adopted from Fig. 3 (a) in Ref. [4], are also shown for comparison. Figure 5 demonstrates excellent agreement between the proposed model and the majority of the experimental points. Although the uncertainties in several parameters would allow the theoretical curves to be tuned more closely to the experimental data, such fine-tuning is not meaningful given the approximate nature of the model and the scatter and error bars in the experimental data themselves.

Turning to analytical predictions, we note that Eq. (48) describes a linear dependence of the LIPSS wavenumber on the pump magnitude, consistent with the observation that the LIPSS period in the periphery of the irradiated spot was 20–40% larger than in its center [4]. The exact distance from the center is not specified in Ref. [4], but our estimates suggest that the local fluence in the periphery was 1.5–2 times lower than at the center. Also, Eq. (48) predicts a sublinear growth of the LIPSS wavenumber with the pump frequency ($\sim\omega^{4/5}$), in qualitative agreement with Ref. [4], where the LIPSS period increased with the pump wavelength, but the observed dependence was faster than linear. The latter discrepancy may be attributed to the decrease of the optimal LIPSS period with increasing relief depth, as discussed in the next section. However, the experimental dependencies in Ref. [4] relies on only a few data points, which makes an accurate extraction of the scaling law difficult.

Overall, we wish to emphasize that the resulting Eqs. (47) and (48) are remarkably robust against uncertainties in most of the material parameters – such as the collision frequency $\nu$, the effective bandgap $W_g$, the ionization probability per unit length $\alpha_0$, and the saturated drift velocity $v_d$ – since it depends on them either as a fifth root or logarithmically. This justifies the simplifying assumptions made above and allows for quantitative theoretical predictions despite the scatter in the experimental data on these parameters.

Calculating the absolute value of the increment $\Omega$ at optimal wavenumbers, we find that the instability development time is several nanoseconds for the parameters of Ref. [4] (see also Fig. 4). This means that the duration of a single pulse train (referred to as a "macro-pulse" in Ref. [4]), equal to 4 μs, is indeed sufficient for strong exponential growth of the temperature perturbation from the noise level. This is consistent with the observation that surface structuring occurs even after a single macro-pulse (containing 108 micro-pulses). The main uncertainty in this estimate is the effective collision frequency of electrons in a strong electric field, $\nu$, which is assumed to saturate at $\sim 10^{15}\ s^{-1}$. Nevertheless, for the chosen set of parameters, the increment remains positive for $\nu \gtrsim 10^{14}\ s^{-1}$.

The instability described above should be considered as the initial stage of LIPSS formation under high-repetition-rate pumping, when the temperature perturbation does not relax between pulses, but the surface is still flat. At later stages, when many pulses have acted on the sample, the gradual modification of the surface relief must also be taken into account. Below we show that the relief deepening modifies conditions for the instability development, leading to the growth of the optimal perturbation wavenumber and decrease of the increment.

## IV. THE ROLE OF THE SURFACE CORRUGATION

In the previous Sections we solve the diffraction problem for a periodic modulation of the permittivity near the flat sample surface, see Eqs. (11) and (21). In a real experiment, however, the sample surface is initially not perfectly flat and, moreover, it is gradually modulated by the train of pump pulses. In this Section we analyze how the diffraction on the surface relief could contribute to the total electric field and change the instability development regime. For this we solve the diffraction problem for a sinusoidally corrugated surface and then superimpose the result onto the total electric field perturbation obtained above, because to first order the two diffraction contributions are additive.

Consider a plane electromagnetic wave of frequency $\omega$ (wavenumber in vacuum $k_0 = \omega/c$) incident normally on the surface of a semiconductor occupying the half-space $x < 0$. The surface is corrugated along the $z$ axis as $h(z) = h_0 \cos(k_g z)$, where $k_0 h_0 \ll 1$, and the medium is assumed to be homogeneous, with a constant background permittivity $\varepsilon$. We consider TE-polarization, so that the electric field has a single non-zero component $E_y(x,z)$.

In the zero order in $h_0 k_0$, the field is given by the standard Fresnel solution for a flat interface: the incident wave $E_{inc} e^{-i\omega t + i k_0 x}$, the reflected wave $E_R e^{-i\omega t + i k_0 x}$ in vacuum, and the transmitted wave $E_y(x,t) = E_y^0 e^{-i\omega t + \eta x - i k_x x}$ inside the medium (the same as in Eq. (6)). In what follows we keep only the transmitted wave, since it is the field inside the medium that drives the heating and ionization considered in the main text.

To first order in $h_0 k_0$, the corrugation manifests itself in two ways: (i) the boundary conditions must be imposed on the actual surface $h(x) = h_0 \cos(k_g z)$ rather than on the plane $x = 0$, and (ii) the tangential component of the magnetic field acquires a correction related to the local tilt of the surface. Both effects can be accounted for by expanding the fields near $x = 0$ and keeping terms linear in $k_0 h_0$.

For the electric field, the continuity condition $E_y\,|_{x=h_0\cos(k_g z)} =$ const gives, to first order,

$$\delta E_y\big|_{x=0^+} + h_0 \cos(k_g z)\, \frac{\partial E_y^{(0)}}{\partial x}\bigg|_{x=0^+} = \delta E_y\big|_{x=0^-} + h_0 \cos(k_g z)\, \frac{\partial E_y^{(0)}}{\partial x}\bigg|_{x=0^-}. \qquad (49)$$

For the magnetic field, the continuity of the tangential component $H_\tau$ on the tilted surface yields

$$\delta H_z|_{x=0^+} + h_0 \cos(k_g z) \frac{\partial H_z^{(0)}}{\partial x}\bigg|_{x=0^+} = \delta H_z|_{x=0^-} + h_0 \cos(k_g z) \frac{\partial H_z^{(0)}}{\partial x}\bigg|_{x=0^-}. \qquad (50)$$

The first-order diffracted field has the form of two counter-propagating waves in each half-space,

$$\delta E_y(x > 0) = A^+ e^{\eta_1 x + i k_g z} + A^- e^{\eta_1 x - i k_g z}, \quad (51)$$

$$\delta E_y(x < 0) = B^+ e^{\eta_2 x + i k_g z} + B^- e^{\eta_2 x - i k_g z}, \quad (52)$$

where, similarly to Eqs. (14) and (15), $\eta_2 = \pm\sqrt{-k_0^2 \varepsilon + k_g^2}$ ($Re\,\eta_2 > 0$ and $Im\,\eta_2 < 0$) and $\eta_1 = \pm\sqrt{k_g^2 - k_0^2}$ ($Re\,\eta_1 < 0$ and $Im\,\eta_1 > 0$).

Substituting (51), (52) into the boundary conditions (49), (50) and using $A^+ = A^- \equiv A$, $B^+ = B^- \equiv B$ (the two counter-propagating waves have equal amplitudes at normal incidence), we obtain a linear system for $A$ and $B$. Its solution gives the complex amplitude of the electric field at the interface:

$$\delta E_y \mid_{x=0} = \frac{h_0 k_0^2 (\varepsilon - 1)}{\eta_2 - \eta_1} \cos(k_g z)\, E_y^{(0)}. \quad (53)$$

Here we should again take into account, by analogy with Eqs. (19)–(21), that the magnitude of the total electric field inside the medium is determined by the real part of Eq. (53):

$$\delta\left|E_y^{tot}\right| \cong \left|E_y^{(0)}\right| Re\left[\frac{h_0 k_0^2 (\varepsilon - 1)}{\eta_2 - \eta_1}\right] \cos k_g z\,. \quad (54)$$

The analytical expression (54) is compared with FEM numerical modelling for the case of strong ionization in Fig. 6, where excellent agreement is observed.

Simplifying Eq. (54) for the limit of high $k_g$ and strong medium ionization, relevant to the experiment [4] and considered in the previous Sections, we obtain:

$$\delta\left|E_y^{tot}\right| \cong -\left|E_y^{(0)}\right| h_0 k_0 \frac{k_0^3}{8 k_g^3} \left(\frac{\omega_p^2}{\omega \nu}\right)^2 \cos k_g z\,. \quad (55)$$

From this expression it follows that diffraction on the surface relief also makes negative contribution to the total electric field in the overheated region. It has a similar form as the contribution from diffraction on the permittivity perturbation (Eq. (26)), and thus it can be incorporated into the equations for the medium ionization and heating.

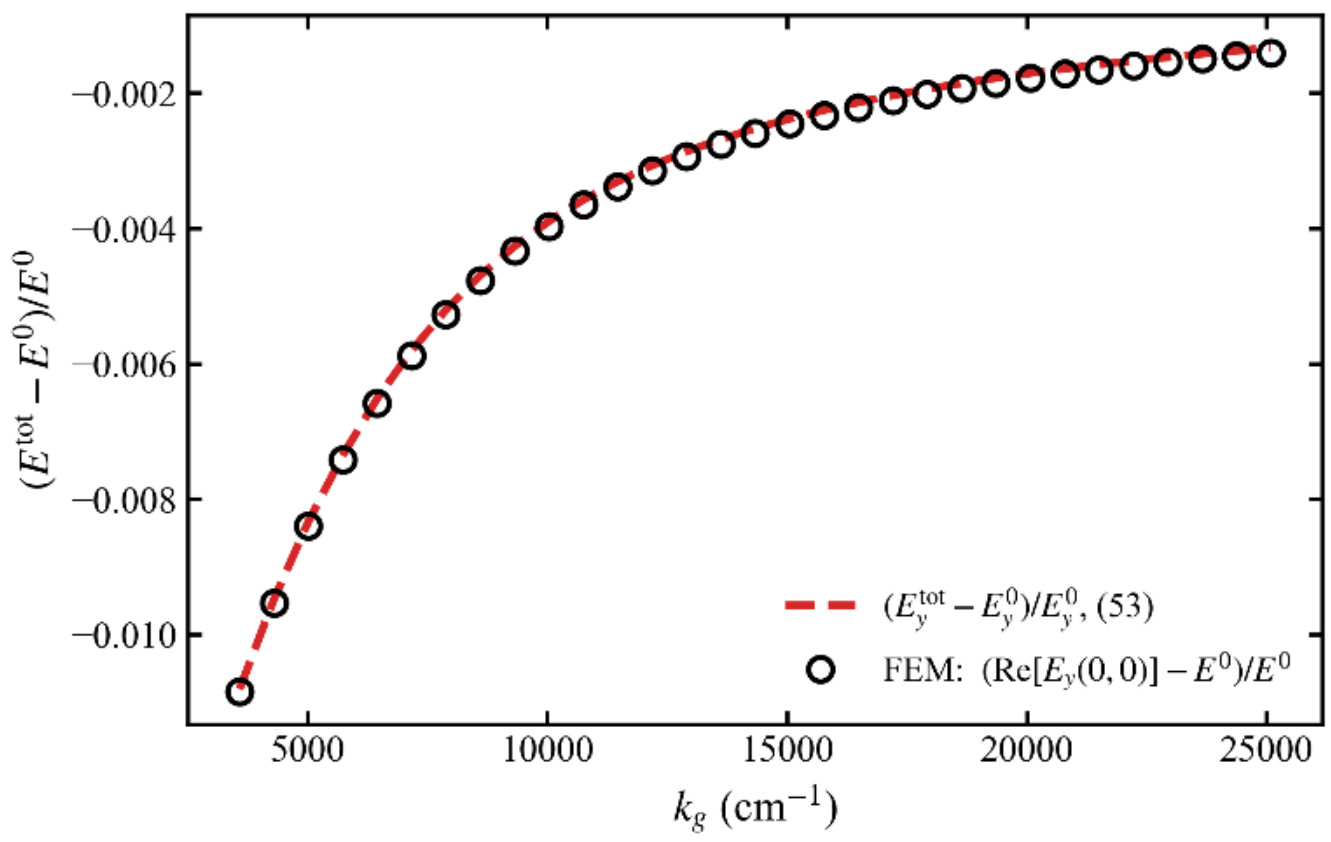

Figure 6. Normalized perturbation of electric field near the semiconductor surface, caused by diffraction on the surface relief, versus the grating wavenumber. The case of strong background medium ionization is considered ($\omega_p^2/\omega\nu \gg \varepsilon_0$) and $k_0 = 10^3\ cm^{-1}$, similarly to Fig. 2. The result of finite-element modelling of the diffraction problem is shown by circles, and the analytical curve given by Eq. (54) is shown by dashed red line.

Let us first consider the grating depth $h_0$ as an independent small parameter, since it does not change significantly during the action of a few pump pulses. Following the logic of Sections II and III, we include the new diffraction contribution (55) in on the medium ionization dynamics, rewriting Eq. (37):

$$\delta n(t) = \delta n_0 \frac{n(t)}{n_0}\left[1 - \frac{1}{72\sqrt{2}}\frac{k_0^3}{k_g^3}\frac{b}{E_{inc}}\frac{\omega_p^4(t)}{\omega^2\nu^2}\right] - n(t)\frac{b}{5E_{inc}}\left[h_0 k_0 \frac{k_0^3}{8k_g^3}\left(\frac{\omega_p^2}{\omega\nu}\right)^{\frac{3}{2}}\right], \tag{56}$$

and Eq. (46) for the medium heating:

$$\frac{\overline{d\delta T_{lat}}}{dt} = \Omega(k_g)\overline{\delta T_{lat}} - \frac{\tau_p}{\tau_d}\frac{e v_d E_{inc}}{12k} h_0 k_0 \frac{k_0^3}{k_g^3}\frac{\omega_p^5}{\omega_a^2(\omega\nu)^{\frac{3}{2}}}, \tag{57}$$

As one can see from Eq. (57), the presence of surface relief changes the conditions of the instability development and, in particular, decreases the increment. As long as we have not introduced a relation between the surface shape and the medium heating, this equation contains two independent parameters: $h_0$ and the magnitude of the temperature perturbation $\delta T_{lat}^0$.

From Eq. (57) we can estimate the shift of the optimal perturbation wavenumber $k_g^*$, considering the linear stage of the instability development, when $\overline{\delta T_{lat}} \cong \delta T_{lat}^0(1+\Omega t)$. In this case, maximizing the increase rate of $\overline{\delta T_{lat}}$ with respect to $k_g$, we find, instead of Eq. (47):

$$k_g^* = \left(\frac{k_0^3}{D_{lat}\tau_d}\right)^{1/5} \times \left(\frac{W_g v_d e E_{inc}\tau_p}{36\sqrt{2}k^2 T_{lat}^2}\frac{\omega_p^4}{\nu\omega\omega_a^2}\left(1 + \frac{b}{4E_{inc}}\frac{\omega_p}{\sqrt{\nu\omega}} + \frac{9h_0 k_0}{\sqrt{2}\delta T_{lat}^0}\frac{kT_{lat}^2}{W_g}\right)\right)^{1/5}. \tag{58}$$

From here it follows that the increase of the grating depth $h_0$ leads to the growth of the optimal perturbation wavenumber and, consequently, to the decrease of the LIPSS spatial period, which coincides qualitatively with the experimental observations from Ref. [4].

The detailed consideration of the surface-shape modification lies far beyond the scope of the present paper. Nevertheless, we can tentatively relate the growth of $h_0$ to the temperature perturbation and obtain quantitative estimates, which should be treated as a working hypothesis rather than as a rigorous result. During the interaction, the semiconductor is strongly heated, and the most heated part of the boundary apparently melts. At the same time, the inhomogeneous heating should give rise to a modulated thermal expansion of the overheated surface layer of the sample. We shall assume that the change of the surface shape after a single pulse is governed by the local thermal expansion of the overheated layer. In other words, we assume that after expansion the material somehow retains its shape, either because of plastic deformation or because the molten part of the material solidifies in its new form.

In the framework of our model, the depth of the layer with the modulated temperature, $l_\delta$, is equal to the penetration depth of the diffraction fields, i.e., $l_\delta = \eta_2^{-1}$, while $\eta_2 \cong k_g$ at high $k_g$ (thus, the layer thickness is about 0.4–1 μm for the considered range of $k_g$). Denoting the thermal expansion coefficient of the material as $\beta$, we can express the modulation of the thermal expansion of this layer $\Delta h_0$ per single pump pulse:

$$\Delta h_0 = l_\delta \beta \delta T_{lat}^0. \tag{59}$$

Thus, after the action of $N_p$ pulses the estimated grating depth is given by

$$h_0 = N_p l_\delta \beta \delta T_{lat}^0, \tag{60}$$

which leads to the following correction to the optimal LIPSS wavenumber:

$$k_g^* = \left(\frac{k_0^3}{D_{lat}\tau_d}\right)^{1/5} \times \left(\frac{W_g v_d e E_{inc}\tau_p}{36\sqrt{2}k^2 T_{lat}^2}\frac{\omega_p^4}{\nu\omega\omega_a^2}\left(1+\frac{b}{4E_{inc}}\frac{\omega_p}{\sqrt{\nu\omega}}+\frac{9}{\sqrt{2}}\frac{k_0}{k_g^*}N_p\beta T_{lat}\frac{kT_{lat}}{W_g}\right)\right)^{1/5}. \quad (61)$$

Choosing $\beta = 4\times10^{-6}\ K^{-1}$ (the thermal expansion coefficient of silicon near its melting point), we find the relative increase of $k_g^*$ of about 3% per $N_p = 10^3$ pulses, which agrees in order of magnitude with the data of Ref. [4], where a relative change of 3–10% per $10^3$ micro-pulses was observed.

Finally, we consider the case of large pulse numbers $N_p$ and relatively deep grating, so that the diffraction on thermal perturbations of the dielectric permittivity can be neglected compared with the diffraction on the modulation of the surface shape. Therefore, the temperature perturbation remains in the equations only as a source of the corrugation depth growth, in accordance with Eq. (60). In this case Eq. (57), combined with the thermal model of $h_0$ growth (Eq. (60)), gives the following instability increment:

$$\widetilde{\Omega}(k_g) = \frac{\tau_p}{\tau_d}\frac{W_g v_d e E_{inc}}{3k^2T_{lat}^2}\frac{\omega_p\sqrt{\nu\omega}}{\omega_a^2} - \frac{\tau_p}{\tau_d}\frac{k_0^4}{k_g^4}\frac{e v_d E_{inc}}{12k}N_p\beta\frac{\omega_p^5}{\omega_a^2(\omega\nu)^{\frac{3}{2}}} - D_{lat}\left(k_g^2+\frac{1}{L^2}\right). \quad (62)$$

The modified increment $\widetilde{\Omega}(k_g)$ reaches its maximal value in the following point:

$$\tilde{k}_g^* = \frac{E_{inc}}{b}\left(\frac{k_0^4}{D_{lat}\tau_d}\right)^{\frac{1}{6}}\left(\frac{16}{3}N_p\beta\frac{ebv_d\tau_p}{k}\frac{\omega\nu}{\omega_a^2}\right)^{\frac{1}{6}}\left(\ln\left(\frac{v_d\alpha_0\tau_p}{2}\right)\right)^{5/6}. \quad (63)$$

Calculating $\tilde{k}_g^*$ for the same set of parameters, as used after Eq. (48) in Section III, we find the optimal LIPSS period of 2.8 μm at $\omega = 5\ THz$ for $N_p = 10^3$, which is in excellent agreement with Ref. [4]. Also, Eq. (63) predicts the LIPSS spatial period to be scaled with the pulse number as $\lambda_g^* \propto N_p^{-1/6}$, which is in qualitative agreement with Ref. [4], where, however, the slower power fitting was found ($\lambda_g^* \propto N_p^{-\vartheta}$, $\vartheta = 0.065 - 0.1$). But, as previously, the extraction of an accurate scaling law is complicated by the rather large experimental uncertainties.

It should be emphasized that Eq. (63) describes the adiabatic limit, in which the relief grows slowly compared with the instability increment, so that the dominant mode follows the shift of the resonance caused by the increasing corrugation depth. In the regime considered here this requirement is satisfied, since $h_0$ changes by only a few percent per macro-pulse, whereas $\Omega\tau_{macro} \gg 1$.

Also note that the two limiting expressions for the optimal wavenumber, Eqs. (47) and (63), are not independent results but two limits of a single resonance condition. In both cases, the optimal wavenumber is selected by the balance between the diffractive weakening of the pump field and the thermal diffusion that smooths out the temperature perturbation. The only difference is the nature of the inhomogeneity on which the pump field diffracts: at $h_0k_0 \to 0$ the dominant contribution comes from the modulation of the permittivity, whereas at large $h_0k_0$ it comes from the corrugated surface. Since both inhomogeneities are small to first order, their contributions add linearly, and the resonance condition reduces to a single equation, containing both terms.

At the initial stage, when the surface is nearly flat, the optimal wavenumber is given by Eq. (47); as the relief grows with the number of pulses, the corrugation term shifts the resonance toward higher $k_g$, and in the limit of deep corrugation the optimal wavenumber is described by Eq. (63). The experimentally observed decrease of the LIPSS period with the number of pulses [4] can therefore be naturally interpreted as a manifestation of this continuous transition. However, we remind that the simple thermal-expansion model used here to relate $h_0$ to the temperature perturbation is only a simplified framework for a preliminary analysis of the general trends rather than a rigorous description of the surface modification.

Above we assumed thermal equilibrium between pump pulses, which is justified for the 37 ns inter-pulse interval of Ref. [4]. We also assumed that the nature of the inter-pulse is simply the residual inhomogeneous heating of the sample surface. Below we discuss limitations of our model using the experiment [5] as an example, where both of these assumptions are invalid.

### V. LIMITS OF APPLICABILITY: EXTREME REPETITION RATES

The theory developed in Sections II–IV relies on two central assumptions: (i) the periodic perturbation that seeds the instability is of thermal origin, i.e. it is a perturbation of the lattice temperature that modulates the equilibrium carrier density via Eq. (38), and (ii) the material response between pump pulses is governed by thermal diffusion, so that the feedback loop is closed through the residual temperature perturbation $\delta T_{lat}$ at the moment of the next pulse arrival. Under these assumptions the optimal wavenumber is given by Eqs. (47) or (64) and, for the parameters of Ref. [4], it reproduces the observed LIPSS period quantitatively.

It is instructive to examine to what extent the same approach can be applied to the experiment of Ref. [5], where LIPSS with a spatial period of about 1 μm (1/10 of the incident wavelength) were observed on silicon under irradiation by mid-IR femtosecond pulses at a repetition rate of 2.8 GHz. The formal estimate for the parameters of Ref. [5] (photon energy well below the bandgap, $\tau_p \simeq 0.5$ ps, $\tau_d \simeq 0.35$ ns, $E_{inc} \simeq 1.6\ \mathrm{MV/cm}$, melting-level lattice temperature) according to Eq. (47) yields a LIPSS period of about 4 μm, which is larger, but still of the same order as the measured one. The limit of high grating depth (Eq. (63)) gives even better result, predicting the LIPSS period of about 2 μm (for the number of pulses $N_p \cong 8 \times 10^4$ used in the work [5]). However, this apparent agreement seems to be misleading, and a closer inspection reveals that the physics of the two experiments is fundamentally different.

First, in the regime of Ref. [5] the inter-pulse delay is comparable to or shorter than the characteristic Auger recombination time, so that the free-carrier density does not return to its thermal-equilibrium value between pulses but instead accumulates over the pulse train. Indeed, in [5] the inter-pulse delay was 0.35 ns, whereas the characteristic Auger relaxation time – even at a carrier density of $10^{20}\ cm^{-3}$ – is about 0.5 ns. Thus, to correctly describe the results of this work, one must explicitly account for both the accumulation of carriers under the action of many pulses and the Auger recombination itself. So that, Eq. (37) for the averaged carriers' concentration takes the following form:

$$\overline{\frac{dn}{dt}} = \frac{\tau_p}{\tau_d} \bar{n} v_d \alpha_0 \exp\left(-\frac{b}{\left|E_y^{tot}\right|}\right) - \Gamma(\bar{n}^3 - n_0^3), \tag{64}$$

where, as previously, $n_0$ is the equilibrium carriers' concentration at current temperature of the medium (see eq. (38)). Consequently, the perturbation of the carrier density that seeds the instability is not simply proportional to the temperature perturbation through the Boltzmann factor (39).

Second, and more importantly, the material modification observed in Ref. [5] is not a purely thermal relief. The TEM and XRD analysis of Ref. [5] demonstrates that the LIPSS produced by the mid-IR FEL contain dislocation-like defects and clusters of point defects extending to a depth of about 1 μm, while the orientation of the crystal lattice as a whole remains essentially undistorted (and the residual strain is below the detection limit). This is in stark contrast to the LIPSS produced by near-IR femtosecond pulses in the same work, which remain crystalline but exhibit a measurable residual tensile strain. The presence of a high density of structural defects in the mid-IR case indicates that the periodic modification is not a smooth temperature grating that could be described by a continuous thermal-diffusion equation, but rather a pattern of localized, irreversible lattice damage. Such a pattern is formed when the accumulated carrier density and the associated local fields exceed the threshold for defect generation, and its spatial period is determined by the interplay between carrier accumulation, Auger recombination, ambipolar diffusion and the screening of the pump field – not by the balance between thermal instability and heat diffusion that determines $k_g^*$ in Eqs. (47) and (48). Constructing a tractable framework that couples defect kinetics to the carrier, lattice-temperature and permittivity equations is therefore an important problem for future research, but it lies beyond the scope of the present work.

For these reasons, we do not attempt to reproduce the precise LIPSS period of Ref. [5] within the present theory. We nevertheless note that the experiments [4] and [5] are not unrelated: in both cases the structuring is seeded by a periodic modulation of the free-carrier density, and in both cases the pump field is screened by the emerging electron-hole plasma. For example, the depth of the modified silicon layer observed in Ref. [5] (about 1 $\mu$m) is comparable to the skin depth in the strong-ionization regime, which is consistent with this picture. The difference lies in the nature of the slow variable that carries the memory of the previous pulses – the lattice temperature in the THz-FEL experiment of Ref. [4], and the accumulated carrier and defect density in the mid-IR FEL experiment of Ref. [5].

## CONCLUSION

We have developed a theory of deeply subwavelength electromagnetic structuring of semiconductors driven by a thermal-ionization instability that develops under the action of a train of intense low-frequency pulses. The proposed model is based on the competition between two physical effects: the medium heating (supported by the impact ionization and decreased by the heat diffusion) and the pumping pulse diffraction. At high repetition rates the temperature perturbation does not relax between pulses, and the competition between these two feedbacks selects an optimal perturbation wavenumber at which the instability increment is maximal. This wavenumber, given analytically by Eq. (47), determines the initial spatial period of LIPSS, while the subsequent thermal deformation gradually shifts this resonance toward higher wavenumbers: Eq. (61) applies at relatively small numbers of pump pulses, and Eq. (63) in the limit of large grating depth (and large numbers of pump pulses).

The main quantitative result of the paper is that, for the characteristics of silicon and of the THz-FEL experiment of Ref. [4], the predicted LIPSS periods are in very good agreement with the measured values, and this prediction contains no fitting parameters. The uncertainty in the material constants at high temperatures affects the predicted period only weakly since the resulting expressions are robust with respect to variations of the most of these parameters. This is the main reason why the theory can be compared quantitatively with experiment despite the scatter in the literature data on hot-carrier transport in silicon.

The theory also reproduces the observed qualitative trends, predicting a linear growth of the optimal wavenumber with pump amplitude, consistent with the reduction of the LIPSS period towards the center of the irradiated spot, and a sublinear growth with pump frequency. The mechanism predicts stripes oriented parallel to the pump polarization, in agreement with Refs. [4, 5] and in contrast to classical SPP-based models, which predict the perpendicular orientation and periods close to the pump wavelength. The estimated instability growth time at the optimal wavenumber is several nanoseconds, much shorter than the duration of a single macro-pulse, which is consistent with the experimental observation that structuring occurs already after a single macro-pulse. The increment remains positive for collision frequencies down to $\nu \simeq 10^{14}\ s^{-1}$, so the mechanism is not fine-tuned to a particular set of parameters, and the theory predicts a smooth onset of structuring rather than a sharp threshold.

We have also explicitly discussed the limits of applicability of the theory in Section V. In the regime of extremely high repetition rates, corresponding to the mid-IR FEL experiment of Ref. [5], the inter-pulse delay is comparable to the Auger recombination time and the material modification is accompanied by the generation of structural defects. In this regime the slow variable that carries the memory of the previous pulses is not the lattice temperature but the accumulated carrier and defect density, and the thermal-ionization instability considered here is apparently not the dominant mechanism. A quantitative description of this regime requires a framework that couples defect kinetics to the carrier, lattice-temperature and permittivity equations – a problem that lies beyond the scope of the present work but that is clearly of interest for future research.

Finally, the predicted scaling of the LIPSS period with pump intensity, pulse duration, repetition rate and frequency provides a set of experimentally testable predictions that can be used to validate the thermal-ionization mechanism. Although the numerical estimates presented here are based on silicon, the theory itself is formulated for a generic semiconductor and does not rely on material-specific features beyond the Drude response and the existence of impact ionization. We therefore expect the mechanism to be relevant to a broad class of semiconductors and dielectrics irradiated by intense low-frequency pulses at high repetition rates.

**Acknowledgements**

The authors are grateful to M. D. Tokman for fruitful discussions.

**Funding**

The work was supported by the Russian Science Foundation, project No. 25-12-00353. The numerical simulations of the diffraction process (Sections I and IV) were carried out by E. Smolina within the framework of the state assignment of the IAP RAS, project FFUF-2026-0003.

**Conflict of interests**

The authors declare no conflict of interest.

**Data availability statement**

The data that support the findings of this study (Sections II and IV) were generated by numerical simulations. The source code and parameters used to generate the simulations are publicly available [19].